\documentclass[prl,amsmath,amssymb,superscriptaddress, twocolumn ]{revtex4-1}
\usepackage{graphicx}
\usepackage{epsfig}
\usepackage{dcolumn}
\usepackage{bm}
\usepackage{rotating}
\usepackage{multirow}
\usepackage[table]{xcolor}
\usepackage[english]{babel}

\begin{document}
\widetext 

\title{ 
Photoemission Signatures of Non-Equilibrium Carrier Dynamics from First Principles
}

\author{Fabio Caruso}
\affiliation{Institut f\"ur Physik and IRIS Adlershof, Humboldt-Universit\"at zu Berlin, Berlin, Germany}  
\author{Dino Novko}
\affiliation{Center of Excellence for Advanced Materials and Sensing Devices, Institute of Physics, Zagreb, Croatia}
\affiliation{
Donostia International Physics Center (DIPC), Donostia-San Sebasti\'{a}n, Spain}
\author{Claudia Draxl}
\affiliation{Institut f\"ur Physik and IRIS Adlershof, Humboldt-Universit\"at zu Berlin, Berlin, Germany}  

\begin{abstract}
{
Time- and angle-resolved photoemission spectroscopy (tr-ARPES) constitutes a 
powerful tool to inspect the dynamics and thermalization of hot carriers. 
The identification of the processes that drive the dynamics, however, is 
challenging even for the simplest systems owing to the coexistence of 
several relaxation mechanisms.  
Here, we devise a Green's function formalism for predicting 
the tr-ARPES spectral function and establish the origin of 
carrier thermalization entirely from first principles. 
The predictive power of this approach is demonstrated by 
an excellent agreement with experiments for graphene over 
time scales ranging from a few tens of femtoseconds up to several
picoseconds. Our work provides compelling evidence 
of a non-equilibrium dynamics dominated by the establishment 
of a hot-phonon regime.}
\end{abstract}

\maketitle

With the advent of ultra-short light sources, 
time- and angle-resolved photoemission spectroscopy 
(tr-ARPES) has emerged as a powerful tool to investigate 
non-equilibrium many-body phenomena with sub-picosecond resolution \cite{RevModPhys.86.959}, 
providing a new momentum to the investigation of dissipation in photo-excited matter.
The time dependence of the transient photoemission intensity 
encodes information on the relaxation mechanisms of 
excited electronic and vibronic states, and it provides 
important information on the ultrafast dynamics of hot carriers.  

{
Overall, the non-equilibrium dynamics of the 
hot carriers in a pump-probe experiment can
be thought of as a three-step process: 
(i) photoexcitation; 
(ii) thermalization of the electrons; 
(iii) energy transfer to the lattice.  
The interaction with a pump pulse in (i) brings the system 
to an initial non-equilibrium regime characterized 
by population inversion, where carriers have 
been promoted from the valence to the conduction bands. 
The electron-electron scattering and 
Auger processes in step (ii) drive the system to a thermalized 
regime, with the electronic occupations well described
by a high-temperature Fermi-Dirac function \cite{bib:gierz13,Gierz/2014,WangPRL12}.
Finally in step (iii), hot carriers dissipate their energy
through the interaction with the lattice, for example, via the 
scattering with acoustic and optical phonons or impurities.
The identification of the scattering processes that drives
dissipation in step  (iii) is of pivotal importance to 
engineer longer hot-carrier lifetimes and propagation lengths, however,  
this challenge is complicated by the coexistence of several competing decay channels.
} 

For graphene, for example, 
 it has been widely debated whether the cooling of the electrons 
and the establishment of thermal equilibrium with the lattice
is governed by {\it supercollisions} \cite{PhysRevLett.109.106602,bib:johannsen13,bib:johannsen15}, 
namely, impurity-assisted scattering to acoustic phonons \cite{bib:SC/Nat13}, 
or rather by  a {\it hot-phonon} regime \cite{Bauer/PRB15}, a 
scenario in which electrons scatter more favorably 
with optical phonons\,\cite{bib:kampfrath05,bib:butscher07}, leading to a higher effective 
temperature for these vibrational modes \cite{bib:allen87}.

First-principles calculations of electron-electron \cite{Hedin1965} and electron-phonon \cite{Ziman:100360,Grimvall:111488}
interaction may provide an answer to such questions by determining accurately the 
rate of each scattering mechanism \cite{bib:giustino17}. 
Even though well-established theoretical tools have emerged to identify the 
signatures of many-body interactions in ARPES \cite{FerdiSatellites,guzzo/2011,caruso/2015,Gumhalter/2016/PRB,Verdi2017}, 
their generalization to time-resolved phenomena is not straightforward.
Non-equilibrium Green's functions (NEGF)  \cite{bib:marini13,bib:sangalli15,bib:molinasanchez17,Bonitz:2113019} are a
powerful tool to investigate the time scales immediately 
after photo-excitation (0-30~fs) \cite{bonitz/pssb19,BonitzPRL18}.  
However, the high computational cost \cite{Balzer:1513054} 
of NEGF calculations has thus far hampered their application 
to longer time scales (30--2000~fs) characteristic of the 
thermalization between electron and lattice degrees of freedom. 

In this Letter, we devise an {\it ab-initio} field-theoretic 
scheme, based on many-body theory of electron-phonon coupling  \cite{bib:giustino17}
and rate equations for the electron-lattice thermalization \cite{bib:allen87,bib:lin08,PhysRevLett.122.016806}, 
to predict the tr-ARPES spectral function for systems driven 
out of equilibrium by the interaction with an ultra-short laser pulse. 
{ This approach enables the identification of the leading mechanism that 
drives the hot-carrier relaxation via interaction with the lattice. 
The accuracy of this approach is validated by the excellent agreement 
with experimental data for graphene, for which we report tr-ARPES 
spectral functions for pump-probe delays ranging from a few tens of 
femtoseconds up to several picoseconds. }

To extend the applicability of many-body theory of
electron-phonon coupling to the study of time-dependent
phenomena, we start by approximating the occupation of the electronic 
levels for the out-of-equilibrium state by an ordinary Fermi-Dirac distribution 
with an effective electronic temperature $T_{\rm el}$. 
This assumption greatly simplifies the computational complexity 
of the problem, as it allows one to resort to the equilibrium 
Green's function formalism.
Its validity, however, is limited to time scales 
subsequent to the thermalization of electronic degrees of freedom \cite{bib:gierz13}. 
At shorter timescales, conversely, electrons may be found 
in a non-thermal regime with occupations pronouncedly
different from those provided by a Fermi-Dirac function.  
In this regime, NEGF provide a more suitable framework for the description 
of the non-equilibrium dynamics \cite{bonitz/pssb19}. 
In analogy to the electrons, the non-equilibrium phonon population
is described by a Bose-Einstein distribution. In order to account for 
the anisotropy of the electron-phonon interaction we consider 
two distinct effective temperatures for strongly-coupled (sc) 
and weakly-coupled (wc) phonons, $T_{\rm sc}$, and $T_{\rm wc}$.
This separation is key to allow for the establishment of a hot-phonon 
regime. The separation between weakly and strongly coupled phonons 
can be rigorously addressed by introducing 
a cutoff $\lambda_{\rm c}$ for the mode-resolved electron-phonon 
coupling strength $\lambda_{{\bf q}\nu}$ at equilibrium, such 
that modes with $\lambda_{{\bf q}\nu} > \lambda_{\rm c}$ 
($\lambda_{{\bf q}\nu} < \lambda_{\rm c}$) are described by  $T_{\rm sc}$ ($T_{\rm wc}$).
After interaction with a laser pulse \cite{sup}, 
the system's dynamics is governed by 
the exchange of energy between electrons and phonons 
via electron-phonon coupling \cite{bib:allen87,bib:lin08}, which drives the system back to equilibrium.
Within this picture, the time dependence of the effective temperatures 
$T_{\rm el}$, $T_{\rm sc}$, and $T_{\rm wc}$ fully defines 
the non-equilibrium dynamics of the system, which we 
obtain via the solution of a three-temperature model (TTM) \cite{bib:allen87,bib:perfetti07,bib:lui10,bib:johannsen13}, 
with all parameters determined from first principles.
A more detailed description of this procedure is reported 
in the Supplemental Material \cite{sup}.

Within these approximations, the time-dependent non-equilibrium self-energy 
due to electron-phonon interaction can be expressed as: 
  \begin{align}\label{eq:sigma}
    &\Sigma_{n{\bf k}} (\omega,\Delta t) = \int\!\frac{d{\bf q}}{\Omega_{\rm BZ}}\sum_{m\nu}
       |g_{mn\nu}({\bf k},{\bf q})|^2 \times \\
  & \left[ \frac { n_{{\bf q}\nu}(\Delta t) + f_{m{\bf k+q}} (\Delta t) }
    {\hbar\omega - \tilde\varepsilon_{m{\bf k+q}} + \hbar\omega_{{\bf q}\nu} }
    + \frac { n_{{\bf q}\nu}(\Delta t) + 1 - f_{m{\bf k+q}} (\Delta t) }
    {\hbar\omega - \tilde\varepsilon_{m{\bf k+q}} - \hbar\omega_{{\bf q}\nu} } \right].\nonumber 
  \end{align}
where $g_{mn\nu}({\bf k},{\bf q})$ are electron-phonon coupling matrix elements, 
$\omega_{{\bf q}\nu}$ are the phonon frequencies,
and $\tilde\varepsilon_{n{\bf k}} = \varepsilon_{n{\bf k}} + i\eta$, with
$\varepsilon_{n{\bf k}}$ denoting Bloch energies and $\eta$ a positive infinitesimal. 
The time dependence of the self-energy is encoded in the out-of-equilibrium fermionic 
and bosonic distributions, given by
$f_{n{\bf k}}  = [e^{\varepsilon_{n{\bf k}} / k_{\rm B} T_{\rm el}} + 1 ]^{-1}$
and $n_{\nu{\bf q}} = [e^{\omega_{\nu{\bf q}}    / k_{\rm B} T_{\nu}} - 1 ]^{-1}$, respectively,
where  $T_{\nu}= T_{\rm sc}$ ($T_{\nu}= T_{\rm wc}$) for the strongly (weakly) coupled modes. 
The evaluation of Eq.~\eqref{eq:sigma} gives 
access to the time- and momentum-resolved spectral function via  
\begin{align}\label{eq:A}
A_{\bf k}(\omega, \Delta t) = 
{1}/{\pi} \sum_n {\rm Im}\,[\omega - \varepsilon_{n{\bf k}}-  \Sigma_{n{\bf k}} (\omega, \Delta t)]^{-1}, 
\end{align}
which, within the sudden approximation, is closely related to the tr-ARPES photoelectron current. 
In the following, we proceed to illustrate the application of this approach to the description of the 
spectral signatures of non-equilibrium carrier dynamics 
in tr-ARPES measurements of graphene. 

The electronic ground-state properties have 
been obtained from density-functional theory \cite{hohenbergkohn,kohnsham1965} in the 
local-density approximation. 
The Kohn-Sham equations have been solved within the plane-wave pseudopotential 
method as implemented in {\sc quantum espresso} \cite{bib:qe} on a homogeneous 
40$\times$40$\times$1 Monkhorst-Pack momentum grid.
Phonon frequencies and eigenvectors have been computed from density-functional 
perturbation theory \cite{bib:baroni01} on a 16$\times$16$\times$1 grid.
Electron and phonon energies and electron-phonon 
coupling matrix elements have been interpolated \cite{giustino/prb07} using 
maximally-localized Wannier functions \cite{bib:wannier}. 
The Fan-Migdal self-energy has been calculated with the {\sc epw} code \cite{bib:epw}, and 
the integral over the phonon momentum  in Eq.~\eqref{eq:sigma} has 
been evaluated on a 400$\times$400$\times$1 grid. 
$p$-type doping due to the interaction with the substrate has been accounted for 
through a rigid shift of the Fermi level by 0.2~eV below the Dirac point, 
which corresponds to a carrier concentration of around $8\times10^{12}$~cm$^{-2}$.
The spectral function has been multiplied by the Fermi-Dirac 
distribution and convoluted with a Gaussian function of 130~meV 
and 0.15~\AA$^{-1}$ widths to account for the finite experimental 
resolution for energy and momentum, respectively \cite{bib:gierz13}. 

\begin{figure}[!t]
\includegraphics[width=0.48\textwidth]{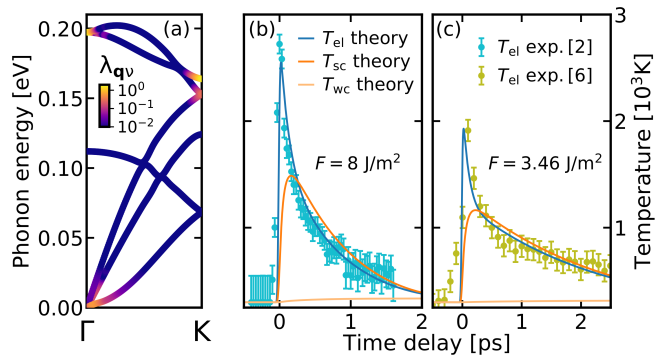}
\caption{\label{fig:fig1}(a) Phonon dispersion and 
mode- and momentum-resolved electron-phonon coupling 
strength $\lambda_{\mathbf{q}\nu}$ along the 
$\Gamma-\mathrm{K}$ path in the Brillouin zone of graphene. 
Time dependence of the effective temperatures of electrons $T_{\rm el}$, 
strongly coupled optical phonons $T_{\rm sc}$, and the remaining lattice vibrations $T_{\rm wc}$, 
as obtained from the TTM for pump fluences $F=8$\,J/m$^2$ (b) and $F=3.46$\,J/m$^2$ (c). 
The experimental electron temperatures (dots) 
are adapted from~\cite{bib:gierz13,bib:johannsen13}. 
}
\end{figure}

The electron-phonon coupling strength $\lambda_{{\bf q}\nu}$, superimposed as 
a color code to its phonon dispersion 
in Fig.~\ref{fig:fig1}~(a), 
is dominated by the electronic coupling to the optical $E_{2g}$ 
and $A_1'$ phonons in proximity to the 
$\Gamma$ and K points in the Brillouin zone, where $\lambda_{{\bf q}\nu}\sim 1$.
This suggests the disentanglement of the $E_{2g}$ and $A_1'$ modes 
from other lattice vibrations, and thus a description of their thermal 
population via $T_{\rm sc}$, whereas $T_{\rm wc}$ is used for the remaining modes. 
This peculiar anisotropy stems from the limited 
phase space available for electron-phonon scattering\,\citep{bib:piscanec04}, 
which is characteristic for multi-valley semiconductors and Dirac metals.
The preferential coupling to the $E_{2g}$ and $A_1'$ phonons
makes them the most likely dissipation mechanism for photo-excited carriers, 
a scenario that may lead to the emergence of hot phonons \cite{bib:wang10}, 
as suggested by recent time-resolved Raman spectroscopy measurements \cite{bib:chae10,bib:berciaud10,bib:wu12,bib:ferrante18}.

\begin{figure*} 
   \includegraphics[width=0.98\textwidth]{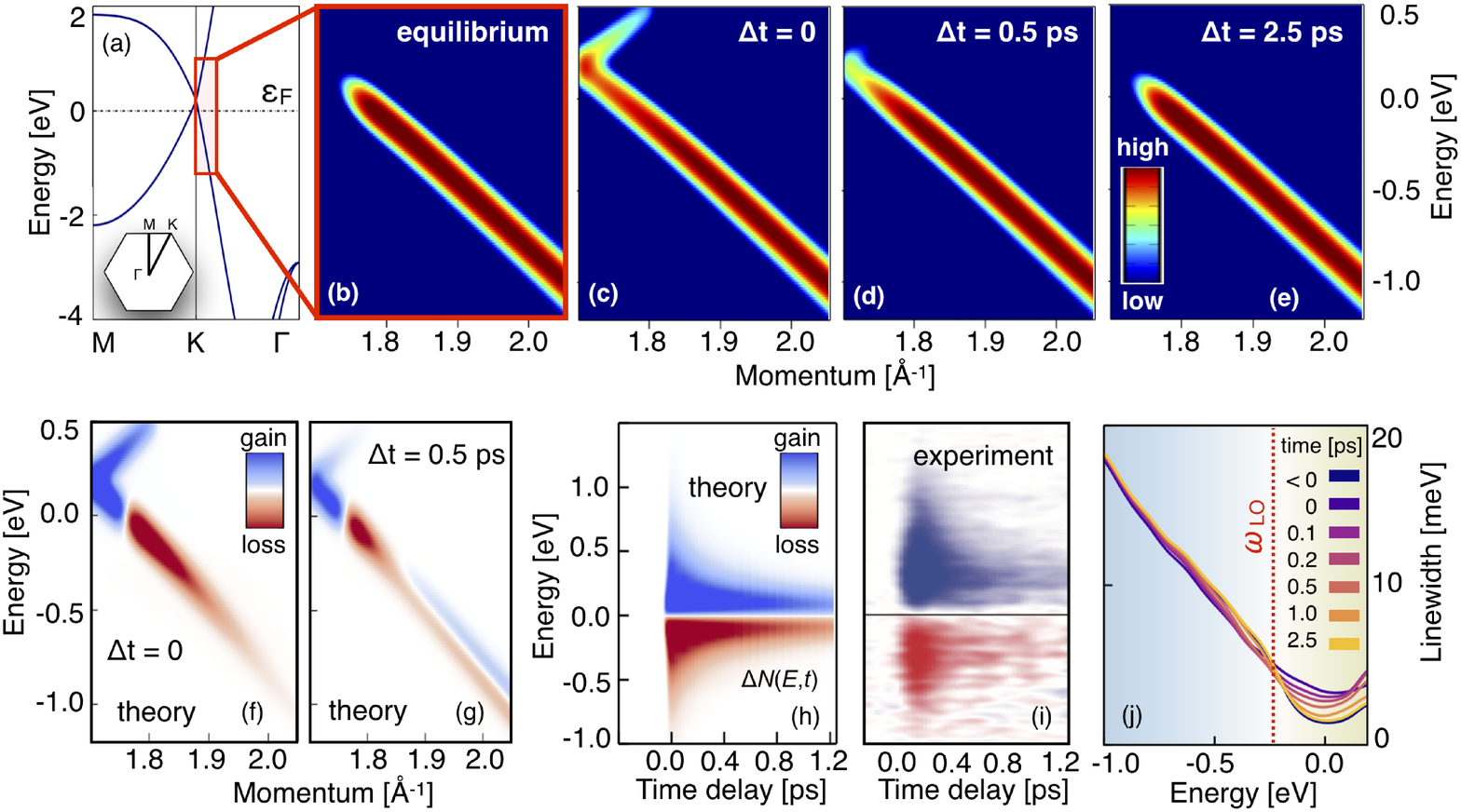} 
\caption{
(a)   Electron band structure of graphene along the 
          high-symmetry line M-K-$\Gamma$ in the Brillouin zone (inset) obtained 
          from density-functional theory. Energies are referenced 
          to the Fermi level (dotted line), which is shifted 0.2~eV below 
          the Dirac point to account for $p$-type doping resulting from the  
          interaction with the substrate. 
(b-e) Calculated tr-ARPES intensity for crystal 
          momenta close to the Dirac point along
          K-$\Gamma$ 
          (red box in panel a) for different time delays between pump and probe beams: 
          before ($\Delta t<0$) and after ($\Delta t=0$) pump, 
          during ($\Delta t=0.5$~ps) and after ($\Delta t=2.5$~ps) the carrier thermalization. 
Calculated differential tr-ARPES intensity relative to equilibrium for
$\Delta t=0$ (f) and 0.5~ps (g). 
Differential pump-probe signal $\Delta N(E,t)$ (integrated over momentum) as obtained from first-principles calculations
(h) and from tr-ARPES measurements (i), reproduced from Ref.~\cite{bib:gierz13}.
(j) Electron linewidths due to electron-phonon interaction at various time delays.
}
\label{fig:spec}
\end{figure*}

Figures~\ref{fig:fig1}~(b) and (c) illustrate 
the time dependence of the electronic and vibrational 
effective temperatures obtained from the 
solution of the TTM, where we considered 
interaction with 30~fs long Gaussian light pulses of 
fluences $F=8$ and $3.46$~J/m$^2$, respectively. 
Below, time delays are relative to the pump switch-off ($\Delta t =0$). 
The abrupt increase of the electronic temperature 
$T_{\rm el}$ to 2550 and 1930~K, respectively, at $\Delta t=0$ marks the photo-excitation of 
hot-carriers following the interaction with the pump, whereas the 
subsequent decrease stems from the thermalization 
of electrons and phonons due to electron-phonon coupling. 
While the temperature of the strongly-coupled phonons $T_{\rm sc}$ approaches 1500~K, 
reflecting their primary role in the hot-carrier relaxation, the temperature of
the remaining lattice vibrations ${T_{\rm wc}}$ increases only mildly. 
The comparison of our results with the experimental 
electronic temperature $T_{\rm el}$  -- extracted from Refs.~\cite{bib:gierz13,bib:johannsen13} 
by fitting a Fermi-Dirac function to the time-resolved (momentum-integrated) 
photoemission signal -- is shown in Figs.~\ref{fig:fig1}~(b) and (c),
and it reveals a remarkable agreement between experiment and theory 
across the full range of the time delays considered.

Beside validating the TTM for the description of 
out-of-equilibrium carrier dynamics over time scales 
ranging from a few tens of femtoseconds to several picoseconds,  
comparison to experiments suggests that hot-carrier relaxation may 
be accurately described within the hot-phonon picture, 
whereby in-plane optical phonons acquire significantly 
higher effective temperature as compared to other lattice vibrations 
owing to their higher coupling strength to electrons.  
Other works reported that hot-carrier relaxation 
may be dominated by supercollisions \cite{bib:johannsen13}, 
impurity-assisted electronic scattering with acoustic phonons.  
However, it has later been argued that this picture would be incompatible 
with the low impurity concentrations of graphene \cite{Bauer/PRB15}. 

After interaction with the pump, 
the change of fermionic and bosonic effective temperatures 
may influence the electron-phonon coupling and its corresponding signatures in spectroscopy. 
In particular, high values of $T_{\rm el}$ increase the phase space for electron-phonon scattering since, 
more final states are available for electronic transitions and 
the thermal phonon population increases with 
$T_{\rm sc}$ or $T_{\rm wc}$, enhancing the scattering rate. 
The interplay of these two trends underpins the 
time dependence of the electron-phonon interaction. 
To quantitatively investigate the signatures of these mechanisms in tr-ARPES 
we proceed to inspect the fingerprints of non-equilibrium carrier 
dynamics in the time-resolved spectral function. 

The tr-ARPES spectral function of $p$-doped graphene, obtained from 
Eqs.~\eqref{eq:sigma} and \eqref{eq:A} for a pump fluence $F=8$~J/m$^2$, is reported in 
Figs.~\ref{fig:spec}~(b-e) for several time delays.
We consider momenta in the immediate vicinity of the Dirac point, 
corresponding to the boxed area in Fig.~\ref{fig:spec}~(a),  
along the high-symmetry line K-$\Gamma$ in the Brillouin zone.
We consider the system at room temperature 
($T_{\rm el} = T_{\rm sc} = T_{\rm wc} =$~300~K)
before the interaction with the pump ($\Delta t<0$). 
The corresponding spectral function, shown in Fig.~\ref{fig:spec}~(b),  
is in good agreement with previous studies \cite{ParkPRB,ParkNano}.
The quasiparticle peaks exhibit a finite linewidth 
resulting from the finite lifetimes of photo-excited holes. The linewidth 
increases linearly with the electron binding energy,
owing to the larger phase space for electron-phonon scattering \cite{ParkPRL}. 
The linewidth is much smaller (3-20~meV) 
than the characteristic energy resolution of tr-ARPES 
experiments (70-130~meV). 
Right after interaction with the pump ($\Delta t=0$), 
the photo-excitation of the system manifests itself 
through a partial population of electronic states 
above the Fermi energy. 
At a time delay $\Delta t =0.5$~ps [Fig.~\ref{fig:spec}~(c)] the 
population of states above the Fermi level decreases, 
reflecting the partial thermalization of the photo-excited carriers.
Finally, for $\Delta t=2.5$~ps [Fig.~\ref{fig:spec}~(d)] the 
tr-ARPES spectral function is virtually identical to the 
one before the pump,  indicating that the system has returned to equilibrium. 

The relative change in photoemission intensity as a function 
of time delay -- obtained as the difference to the spectral function 
before the pump [Figs.~\ref{fig:spec}~(f-g)] --  provides 
further insight into the spectral fingerprints of the quasiparticle 
dynamics in tr-ARPES. The most prominent change in the 
spectral function results from the enhanced concentration of electrons (holes)
above (below) the Fermi surface, induced by the photo-excitation.
At zero pump-probe delay [Fig.~\ref{fig:spec}~(f)] the spectral intensity gain (blue) 
extends up to 1~eV above the Fermi energy, with a corresponding 
intensity loss (red) for binding energies up to $1$~eV. 
For $\Delta t=~$0.5~ps [Fig.~\ref{fig:spec}~(g)], 
the gain signal is found only up to the Dirac point, 
owing to the decrease of $T_{\rm el}$ at larger pump-probe delays. 
Figure~\ref{fig:spec}~(g) further reveals gain for binding energies 
larger than 0.3~eV. At variance with the features in the 
immediate vicinity of the Fermi energy, these features may not be 
explained in terms of transient electronic temperature, since they 
extend to energies beyond the domain of thermal temperatures. 
Conversely, these features reveal a time-dependent renormalization  
of the quasiparticle energies. 
This finding is compatible with recent observations
in transient optical measurements \cite{Parmigiani11,bib:roberts14} 
of narrowing energy gap between the $\pi$ and $\pi^*$ bands.

This effect can be quantified by inspecting the change in the real part of 
the electron self-energy, 
$\Delta\Sigma = {\rm Re}[\Sigma_{n{\bf k}}(\varepsilon_{n{\bf k}}, \Delta t) 
- \Sigma^{\rm eq}_{n{\bf k}}(\varepsilon_{n{\bf k}})],$ 
where 
$\Sigma^{\rm eq}$ denotes the Fan-Migdal self-energy of the system at equilibrium.
$\Delta\Sigma$ provides information pertaining to the time-dependent change 
of the quasiparticle energies 
due to electron-phonon interaction.
For $\varepsilon_{n{\bf k}}=1$~eV, we obtain $\Delta \Sigma =10$~meV at $\Delta t = 0.5$~ps.
On the other hand, at $\Delta t = 0$~ps, when $T_{\rm sc}$ 
approaches its maximum value, the quasiparticle energies 
are almost unchanged ($\Delta \Sigma <1 $~meV). This suggests
that the energy renormalization (at these lower energies) stems from 
the electronic coupling to acoustic modes. 

To enable the comparison with tr-ARPES measurements 
across the full range of time delays, we evaluate the differential 
momentum-integrated pump-probe signal 
$\Delta N(\omega, \Delta t) = \int d{\bf k} 
[ A_{\bf k}(\omega,\Delta t) -  A^{\rm eq}_{\bf k}(\omega)]$, where 
$A_{\bf k}^{\rm eq}$ denotes the spectral function at equilibrium. 
Our calculations for $\Delta N(\omega, \Delta t)$ are reported in 
Fig.~\ref{fig:spec}~(h), whereas experimental data from \cite{bib:gierz13}
are shown in Fig.~\ref{fig:spec}~(i). 
The comparison with experiment demonstrates good 
agreement across the full range of time delays. 

Finally, we discuss the time dependence of the electron linewidths due to 
electron-phonon interaction, illustrated in Fig.~\ref{fig:spec}~(j).
Close to the Fermi energy, the linewidths exhibit a fast increase 
after the pump, followed by a slow decrease. 
After 2.5~ps, the linewidth coincides with the 
one preceding photo-excitation ($\Delta t<0$), 
marking the return to equilibrium.
This behavior can be ascribed to the time-dependent change of  
the phase space for electron-phonon scattering, 
which is enhanced at high electronic temperatures $T_{\rm el}$. 
At binding energies larger than $\sim180$~meV, on the other hand, 
the onset of strongly-coupled phonons (vertical dotted line)
provides an additional scattering channel that underpins 
the enhancement of linewidth. 
At these binding energies, the linewidth depend only slightly on time.

As an outook, a richer scenario of many-body interactions may emerge in 
compounds characterized by low-energy polar phonons, such as, 
for example, metal-monochalcogenides (e.g., SnSe, SnS)
and transition-metal dichalchogenides (e.g., MoS$_2$, WS$_2$).
These compounds are good candidates for the 
observation of spectral fingerprints of hot-phonon dynamics, 
owing to their large Born effective charges and the 
preferential coupling to few LO modes. 
Furthermore, at variance with graphene, 
the crystal dynamics is characterized by soft lattice vibrations \cite{Caruso/prb/19,Molina2011}
with energies comparable to the room temperature thermal energy (25~meV).
Their population may thus be strongly enhanced by
the increase of the effective lattice temperature \cite{Caruso/prb/19}, a phenomena that may 
underpin a strong increase of electron-phonon interaction upon pump.
This may lead to the emergence of time-dependent signatures of many-body phenomena in 
tr-ARPES, such as, transient satellites and  kinks. 

{In conclusions, we have devised a first-principles scheme
to predict transient many-body phenomena in tr-ARPES for 
systems driven out of equilibrium by the interaction with 
an ultra-short laser pulse. 
Based on this approach, we have shown that carrier relaxation 
in graphene is dominated by the establishment of a hot-phonon regime, 
whereby in-plane optical phonons provide the primary decay path 
for photo-excited carriers.}
The excellent agreement with available experimental 
data substantiate our findings and the validity of 
the hot-phonon picture over time scales ranging between a  
few tens of femtoseconds up to several picoseconds. 
Overall, this work defines a solid strategy to decipher 
the quantum mechanical origin of non-equilibrium 
phenomena in tr-ARPES, paving the way to explore 
the emergence of transient many-body processes 
in spectroscopy. 

\acknowledgements
FC and CD gratefully acknowledge financial support 
from the German Science Foundation (DFG) through 
the Collaborative Research Center HIOS (SFB 951).
DN gratefully acknowledges financial support from the 
European Regional Development Fund for the “Center of 
Excellence for Advanced Materials and Sensing Devices” 
(Grant No. KK.01.1.1.01.0001) and Donostia International
Physics Center (DIPC). Part of the computational resources
were provided by the DIPC computing center.


%

\end{document}